\documentclass[sigconf]{acmart} 
\AtBeginDocument{%
  \providecommand\BibTeX{{%
    \normalfont B\kern-0.5em{\scshape i\kern-0.25em b}\kern-0.8em\TeX}}}

\setcopyright{acmcopyright}
\copyrightyear{2023}
\acmYear{2023}
\setcopyright{acmlicensed}\acmConference[Koli Calling '23]{23rd Koli Calling International Conference on Computing Education Research}{November 13--18, 2023}{Koli, Finland}
\acmBooktitle{23rd Koli Calling International Conference on Computing Education Research (Koli Calling '23), November 13--18, 2023, Koli, Finland}
\acmPrice{15.00}
\acmDOI{10.1145/3631802.3631832}
\acmISBN{979-8-4007-1653-9/23/11}

\usepackage{xcolor}
\usepackage{listings}
\usepackage{graphicx}
\usepackage{xparse}
\usepackage{array}

\def\codeword#1{\texttt{\textcolor{darkgray}{#1}}}
\newcommand{\initspace}{\textcolor{darkgray}{ \texttt{\underline{\hspace{0.3cm}}init\underline{\hspace{0.3cm}} }}}
\newcommand{\init}{\textcolor{darkgray}{\texttt{\underline{\hspace{0.3cm}}init\underline{\hspace{0.3cm}}}}}
\newcommand{\str}{\textcolor{darkgray}{\texttt{\underline{\hspace{0.3cm}}str\underline{\hspace{0.3cm}}}}}
\newcommand{\strspace}{\textcolor{darkgray}{\texttt{\underline{\hspace{0.3cm}}str\underline{\hspace{0.3cm}} }}}
\begin{document}

\title[Understanding the Effects of Using Parsons Problems to Scaffold Code Writing]{Understanding the Effects of Using Parsons Problems to Scaffold Code Writing for Students with Varying CS Self-Efficacy Levels}

 
\author{Xinying Hou}
\orcid{0000-0002-1182-5839}
\affiliation{%
  \institution{University of Michigan}
  \city{Ann Arbor}
  \state{Michigan}
  \country{USA}
}
\email{xyhou@umich.edu}

\author{Barbara J. Ericson}
\orcid{0000-0001-6881-8341}
\affiliation{%
  \institution{University of Michigan}
  \city{Ann Arbor}
  \state{Michigan}
  \country{USA}
}
\email{barbarer@umich.edu}

\author{Xu Wang}
\orcid{}
\affiliation{%
  \institution{University of Michigan}
  \city{Ann Arbor}
  \state{Michigan}
  \country{USA}
}
\email{xwanghci@umich.edu}

\begin{abstract}
 Introductory programming courses aim to teach students to write code independently. However, transitioning from studying worked examples to generating their own code is often difficult and frustrating for students, especially those with lower CS self-efficacy in general. Therefore, we investigated the impact of using Parsons problems as a code-writing scaffold for students with varying levels of CS self-efficacy. Parsons problems are programming tasks where students arrange mixed-up code blocks in the correct order. We conducted a between-subjects study with undergraduate students (N=89) on a topic where students have limited code-writing expertise. Students were randomly assigned to one of two conditions. Students in one condition practiced writing code without any scaffolding, while students in the other condition were provided with scaffolding in the form of an equivalent Parsons problem. We found that, for students with low CS self-efficacy levels, those who received scaffolding achieved significantly higher practice performance and in-practice problem-solving efficiency compared to those without any scaffolding. Furthermore, when given Parsons problems as scaffolding during practice, students with lower CS self-efficacy were more likely to solve them. In addition, students with higher pre-practice knowledge on the topic were more likely to effectively use the Parsons scaffolding. This study provides evidence for the benefits of using Parsons problems to scaffold students' write-code activities. It also has implications for optimizing the Parsons scaffolding experience for students, including providing personalized and adaptive Parsons problems based on the student's current problem-solving status.

\end{abstract}



\maketitle

\section{Introduction}
Commonly used techniques to introduce a new programming topic in college lectures often involve direct instruction and worked example code demonstration \cite{luxton2018introductory}. After
this, students are expected to gain expertise by solving more programming problems independently. However, while students appear to understand the theoretical concepts and examples taught in lectures, transitioning to writing full solutions to new problems remains a huge challenge \cite{ericson2017solving, lister2020cognitive, kreitzberg1974cognitive,sentance2017computing}. As a result, they often fail to overcome the challenge of writing their own code independently, particularly students with low CS self-efficacy, who are less likely to persevere when faced with difficulties \cite{zeldin2000against, wang2022examining}.

To tackle this issue, prior work has investigated different scaffolding approaches to assist students in learning to write code. Scaffolding refers to the assistance given to someone to help them complete a task when they cannot do it independently yet \cite{kim2011scaffolding}. One-on-one tutoring, which can provide the desired scaffolding, has been found to be more effective than traditional classroom instruction with one instructor \cite{bloom19842}. However, as computer science becomes increasingly popular, the number of students in undergraduate CS classes has grown considerably in recent years, which in turn presents serious challenges to offering one-on-one tutoring due to the high cost \cite{guo2015codeopticon}. 

Parsons problems are an increasingly popular type of programming exercise that requires students to place mixed-up code blocks in the correct order \cite{parsons2006parson, ericson2022parsons}. Previous research has shown that Parsons problems generally require less cognitive load from students compared to write-code problems \cite{haynes2021problem} and facilitate greater engagement \cite{ericson2015analysis}. This inspired our work to investigate whether Parsons problems can be used as effective scaffolding when students are writing code for new topics. As a type of completion problem, they have the potential to help students transition from worked examples to conventional write-code programming problems \cite{paas2003cognitive, van2005cognitive}. 

To understand the effectiveness of Parsons problems as a scaffolding technique when students are learning to write code, we conducted a between-subjects classroom experiment. In the experiment, we wanted to understand whether and how students with distinct levels of self-efficacy differ in practice and learning when they received Parsons problems as scaffolding (Parsons scaffolding - PS condition) during write-code practice versus not (NP condition). Furthermore, we performed in-depth analyses to understand students' experiences when interacting with Parsons problems as a scaffolding technique, and the relationship between their interaction and CS self-efficacy levels. We examined the following research questions:
\vspace{-1.8mm}
\begin{itemize}
    \item RQ1.1: Are there differences between conditions in terms of practice performance, problem-solving efficiency, and posttest performance for students with low CS self-efficacy levels?
    \item RQ1.2: Are there differences between conditions in terms of practice performance, problem-solving efficiency, and posttest performance for students with high CS self-efficacy levels?
    \item RQ2: In the Parsons Problems as Scaffolding (PS) condition, how did students with varying CS self-efficacy levels use the Parsons scaffolding?
    \item RQ3: In the Parsons problems as Scaffolding (PS) condition, how did students rate the usefulness of the Parsons scaffolding and why?


\end{itemize}

\section{Related Work}

\subsection{Scaffolding Write-Code Problems}
Scaffolding strategies help students finish a task or build new understanding so that they can perform comparable activities on their own later. By providing desired scaffolding, \citeauthor{bloom19842} demonstrated that one-on-one human tutoring helps students improve their learning by two standard deviations over typical classroom instruction with a single teacher for 30 students \cite{bloom19842}. However, providing this type of support in high student-to-teacher ratio courses, such as introductory CS courses, can be too expensive.

To address this issue, researchers have looked into computer-assisted scaffolding techniques at various stages of code development. One line of research provides scaffolding to restrict the entry difficulty of code-writing to avoid cognitive overload. For instance, \citeauthor{denny2019closer} showed that letting students review and think about problem statements before writing any code had a positive influence on performance \cite{denny2019closer}. Similarly, \citeauthor{garcia2021evaluating} tested design-based Parsons problems which asked students to put strategic plans in order, and found that some students used these problems to try to understand the problem better but that others just used a trial and error approach to solve the problem \cite{garcia2021evaluating}.

Another line of research focuses on providing scaffolding throughout the code-writing process, such as by providing next-step hints in Intelligent Tutoring Systems (ITS). For example, \citeauthor{rivers2017automated} built ITAP, which employs a three-stage process to generate next-step hints for student code submissions \cite{rivers2017automated}. Her initial analysis found that students with hints spent less time practicing but achieved the same learning outcomes. Nevertheless, the inconsistency in the quality of the automated hints is still a problem, affecting students' trust in these systems and future help-seeking behaviors \cite{price2019comparison}. Furthermore, automated hints rarely contain comprehensive guidance, such as examples, that might be leveraged to overcome the "design barriers" experienced by beginner programmers \cite{ko2004six}. As a result, by using low-level hints before starting to program a task, some novices experienced inefficient help-seeking outcomes \cite{marwan2020unproductive}. As we aim to provide scaffolding for students in the early stages of acquiring coding skills on this topic, we would like to implement a more comprehensive scaffolding method during the code-writing process, which is providing students with equivalent Parsons problems alongside the write-code problems. 


\subsection{Existing work on Parsons problems}
In the original design of Parsons problems, \citeauthor{parsons2006parson} provided students with a problem description and a set of drag-and-drop code fragments \cite{parsons2006parson}. Each code fragment was made up of one or more lines, and some of the lines included incorrect code. To complete a problem, students chose the correct code segments and arranged them in the correct order. They reported that the majority of students thought this type of problem was useful for learning \cite{parsons2006parson}. 

Subsequent research has produced a range of Parsons problem varieties, and the difficulty levels of these formats vary depending on the tasks students need to complete and the code blocks presented. For instance, in one-dimensional Parsons problems, code blocks must be organized in the right vertical sequence, while in two-dimensional Parsons problems, the blocks must additionally be appropriately indented \cite{ihantola2011two}. In addition, \citeauthor{weinman2021improving} proposed faded Parsons problems where students must use valid expressions to fill in blanks in the lines of code and rearrange the lines of code to create a proper program \cite{weinman2021improving}. Distractors, code blocks that are not needed in a correct solution, can also be added to make the problems more challenging \cite{ericson2018evaluating}. Distractor blocks typically include syntactic or semantic flaws. There are two types of display for distractor blocks: paired distractors, which contain some indication that students have to pick one of a set, and unpaired distractors, which are randomly mixed in with the correct blocks. Previous research has demonstrated that paired distractors make Parsons problems easier to solve than unpaired distractors \cite{denny2008evaluating}. 

\citeauthor{ericson2019investigating} created two types of adaptation for Parsons problems: intra-problem and inter-problem adaptation \cite{ericson2019investigating}. Intra-problem adaptation reduces the difficulty of the current Parsons problem, and inter-problem adaptation affects the difficulty of the next problem based on students’ performance on the current problem \cite{ericson2018evaluating}. Learners can initiate the intra-problem adaptation by clicking the "Help Me" button after at least three full attempts, and the system will either remove a distractor block or combine two blocks into one. Inter-problem adaptation happens after the system evaluates the learner's performance on the previous Parsons problem. To make the next problem easier, it will remove or pair distractors with the correct blocks. To make it harder, it will add all distractors or randomly mix them with the correct blocks. In this study, we use two-dimensional Parsons problems with intra-problem adaptation and paired distractors.

\subsection{Cognitive Load Theory}
According to cognitive load theory, the human cognitive architecture is made up of numerous memory stores, including a restricted working memory and an unlimited long-term memory \cite{schnotz2007reconsideration}. Working memory is limited in terms of capacity and duration, especially when processing new information. Cognitive load refers to the working memory resources necessary when learning new information \cite{sweller1994some}. Two core cognitive load categories are intrinsic and extrinsic cognitive load \cite{sweller2019cognitive}. Intrinsic cognitive load relates to the material's inherent difficulty, which is mediated by the learner’s prior knowledge \cite{sweller2019cognitive}. Instructional strategies like segmenting and pre-training can be applied to manage intrinsic cognitive load due to overly complex content \cite{mayer2003nine}. Extraneous cognitive load is determined by how the instructional information is delivered and what the learner is expected to perform \cite{sweller2019cognitive}. The extraneous load can be reduced by dealing with typical instructional elements that may cause extraneous load \cite{ccakirouglu2017exploring}.\par

Computer programming is a highly cognitive skill that requires mastering multiple competencies and is recognized as being inherently difficult to learn, making cognitive load theory one of the most relevant theories in computing education research \cite{berssanette2021cognitive}. Previous studies have adopted a wide range of instructional recommendations provided by cognitive load theory to programming learning. Among those effects, the use of worked examples is extremely popular in introductory programming courses \cite{berssanette2021cognitive}. Worked examples demonstrate an expert's comprehensive solution to a problem, allowing students to learn how to solve problems before they can write correct code independently. They are commonly used for novice learners to reduce cognitive load \cite{sweller2010element}. However, in the traditional way of employing programming examples, there are dramatic shifts in cognitive demand when moving directly from fully completed examples to solving a problem by writing code from scratch  \cite{renkl2004fading}. To bridge this gap, one recommended method is to use completion problems with a partial answer before expecting students to complete a full problem \cite{sweller2019cognitive}. Since Parsons problems provide students with the right code blocks but still require them to arrange them in the right order, they fall under the category of completion problems. Therefore, by adding Parsons problems as scaffolding to code writing problems, we expect to implement a smoother transition for students from studying examples to solving write-code problems independently. 
\vspace{-2mm}
\subsection{Self-efficacy in CS Learning}
Self-efficacy describes people's perceptions of their own abilities to complete a task \cite{bandura1977social}. Self-efficacy is important in education as individuals' self-efficacy can influence their willingness to put effort into a task, decisions about future involvement in a task, and attitudes when facing obstacles \cite{bandura1977self}. For example, students with high self-efficacy are usually more enthusiastic about participating in and completing learning tasks than students with low self-efficacy. When facing difficulties, students who believe in their abilities (high self-efficacy) do not avoid difficult tasks but view them as challenges that must be overcome. For example, science learners with a greater degree of self-efficacy will put in more effort on learning activities and will persevere when faced with difficulties, resulting in learning success \cite{britner2006sources, zeldin2000against}.

When it comes to the CS domain, self-efficacy has become among the most studied constructs to understand programming learning outcomes and persistence in computing learning and careers \cite{prather2020we}. Specifically, \citeauthor{wiedenbeck2005factors} discovered that self-efficacy was positively connected with two distinct programming course outcomes: performance in debugging tasks and the overall course grade \cite{wiedenbeck2005factors}. In another study, \citeauthor{lewis2011deciding} interviewed 31 students at two public universities and discovered that one crucial aspect in their decision to major in CS was their perception of their CS ability (self-efficacy in CS) \cite{lewis2011deciding}. Similarly, \citeauthor{miura1987relationship} applied self-efficacy surveys and found students with higher self-efficacy were more likely to enroll in a computer science course in the college \cite{miura1987relationship}.  In this work, we will investigate how students with varying levels of CS self-efficacy used the Parsons scaffolding during practice and whether there are differences between conditions in terms of practice performance, in-practice problem-solving efficiency, and posttest performance for students with various CS self-efficacy levels.

\vspace{-2mm}
\subsection{Using Parsons problem to Scaffold Writing Code}\label{scaffolding-section}
Parsons problems have been explored for both formative assessment (practice) and summative assessment \cite{ericson2022parsons}. Prior studies have provided evidence that most students find Parsons problems engaging \cite{parsons2006parson}, and students can achieve the same level of learning as writing the equivalent code, but with higher learning efficiency \cite{ericson2018evaluating}. \citeauthor{morrison2016subgoals} also reported that Parsons problems were more sensitive than writing code for assessing students' learning gains \cite{morrison2016subgoals}. Parsons problems allow students to demonstrate their understanding of the meaning and sequence of programs, which helps to assess students' knowledge in ways that writing code alone cannot. 

An initial study of Parsons problems as scaffolding explored when, why, and how students apply Parsons problems to scaffold their write-code problems \cite{hou2022using}. The think-aloud study found that students opened the Parsons problem at three different stages: planning, implementing, and debugging a solution. In addition, \citeauthor{hou2022using} identified four distinct ways in which students successfully interacted with Parsons problems to help them complete coding tasks: "scan Parsons problem", "attempt Parsons problem", "solve and replace their code", and "solve and modify their code." \cite{hou2022using}. In "scan Parsons problem", the learner looks at the unsolved Parsons problem for ideas about how to get started or for particular information, but does not attempt to solve it. In "attempt Parsons problem", the learner makes some effort to solve the Parsons problem but does not completely solve it before finishing the write-code problem. In "solve and replace their code", learners solve the Parsons problem and use that solution to replace any code they wrote in the write-code problem. In "solve and modify their code", learners solve the Parsons problem and then modify their solution in the write-code problem.
When experimenting with the learning effectiveness of Parsons problems to scaffold write-code problems, \citeauthor{hou2022using} discovered that students who were given Parsons problems as scaffolding for code writing problems took less time to complete those problems, however, there was no learning improvement from pretest to posttest in either condition, indicating that those students had already mastered the topic \cite{hou2022using}. In contrast to the prior study, we chose a more advanced programming topic to which students had little prior exposure. Our research investigated if Parsons problems can help students bridge the gap between learning from worked examples and solving write-code problems independently.
\vspace{-3mm}
\section{Method} \label{Method}
Our IRB-approved study was conducted in the fall semester of 2022 at a large public research university in the northern United States. All participants were enrolled in a data-oriented programming course, which was the second required Python course for the university's information science majors, though other majors take the course as well. This course covered programming concepts including Python basic data structures (list, tuple, and dict), object-oriented programming concepts (\codeword{classes, objects, and inheritance}), how to debug, how to use unit testing, basic web scraping, regular expressions, HTML, XML, JSON, working with APIs, working with databases, and Matplotlib.

\subsection{System Interface}
\begin{figure}[ht]
    \centering
    \includegraphics[width=\linewidth]{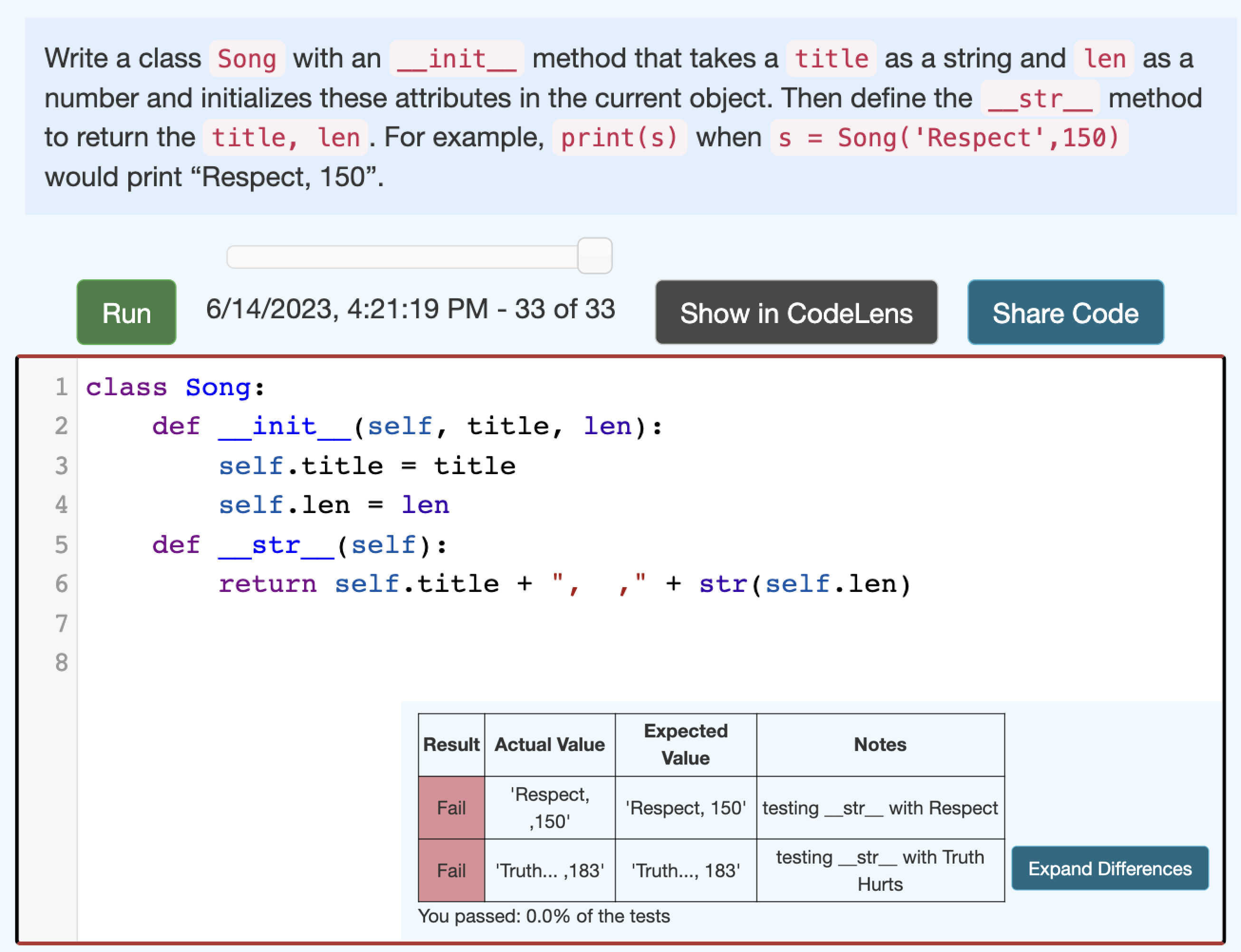}
    \caption{Screenshot of a write-code problem with the unit test results}
    \label{write}
    \vspace{-5mm}
\end{figure}

Runestone allows students to write, execute code, and receive immediate feedback from unit test results (Figure \ref{write}) \cite{ericson2020runestone}. Our study added an equivalent optional Parsons problem to each write-code problem to scaffold students' code-writing practice. When students have difficulty solving a write-code problem independently, they could display, interact with, and work on the equivalent Parsons problem in a preview window (Figure \ref{interface}), but they were still required to solve the write-code problem to earn points. Students could not copy and paste the Parsons solution to the write-code problem, they had to retype it. When students failed to pass all the unit tests after three submission attempts, they would receive a pop-up prompt reminding them that help is available ("Help is Available Using the Toggle Question Selector").

\vspace{-2mm}
\subsection{Participants and Procedure}
\begin{figure}[ht]
    \centering
    \includegraphics[width=\linewidth]{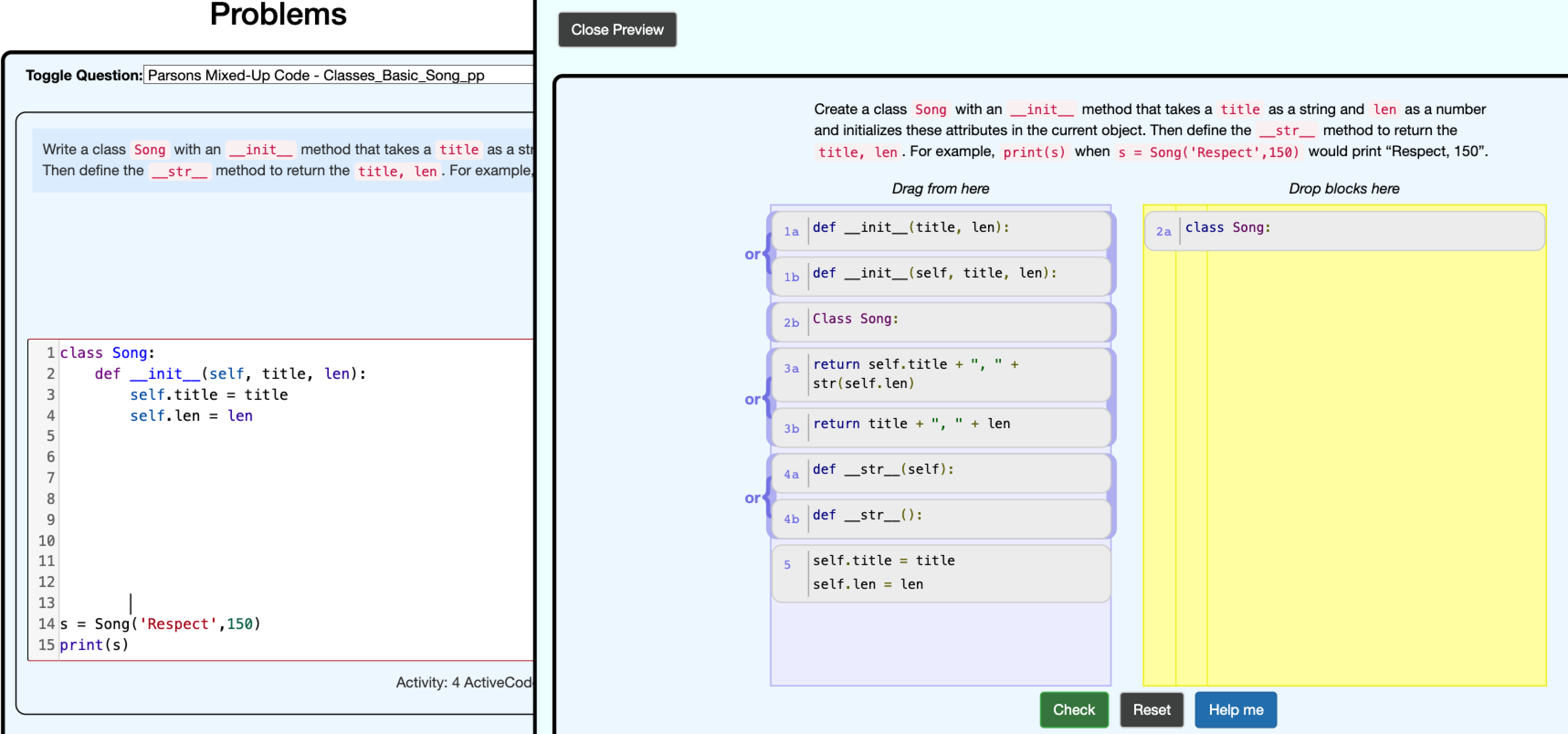} \\
    \caption{Screenshot of using a Parsons problem to scaffold a write-code problem}
    \label{interface}
    \vspace{-5mm}
\end{figure}

The classroom study was conducted during the 80-minute lecture period in week three of the class; students who did not attend the lecture were allowed to finish the study by the end of the day. A total of 134 students participated in this study. Students were randomly assigned to one of two conditions: Parsons-Scaffolding condition (PS) and No-Parsons-Scaffolding condition (NP). Students in the Parsons-Scaffolding (PS) condition received a text-entry write-code interface with an equivalent two-dimensional adaptive Parsons problem as scaffolding (Figure \ref{interface}), while the No-Parsons Scaffolding (NP) group only had the write-code interface (Figure \ref{write}). 

Students first read an introduction to Python \codeword{classes}. It covered three fundamental concepts: defining a new \codeword{class}, constructing new \codeword{objects}, and writing new \codeword{methods}. Each concept included textual instruction, an executable worked example followed by a task where students were instructed to modify the code (Figure \ref{worked_example}). After finishing the introduction to the concepts, students received an introduction to the types of problems in the system: scaffolded code writing or non-scaffolded. Following that, students were asked to complete a programming self-efficacy survey (6 questions) and a self-evaluation on their knowledge about writing code for \codeword{classes} (4 questions). Then, students were given four write-code practice problems in each of the two conditions, with the only difference being that students in PS condition had the equivalent Parsons problems as scaffolding. Students in PS condition were explicitly told that they had to enter code in the write-code area to earn points. We generated a random number between 1 to 10 once the student clicked to start the practice. Based on this number, we assigned them to the NP condition if it was odd and the PS condition if it was even. After each practice question, students in the PS condition were given a question, asking them to rate the perceived usefulness of using a Parsons problem to help them solve the write-code problem on a Likert scale from 1-low to 9-high. Students who did not use the Parsons scaffolding were asked to skip the corresponding survey question. 

Additionally, following the last practice question, PS students were asked to respond to a brief open-ended question explaining their general perceived usefulness of a Parsons problem as scaffolding while writing code. Students in both conditions then finished the posttest. Each section had no time restrictions, allowing students to progress through the materials at their own pace until the end of the day. The final sample contained 89 students who completed all of the materials in order (41 in PS and 48 in NP). Only three students chose to finish the materials outside of class time. They were not outliers after checking their main measures, including practice performance, problem-solving efficiency, and posttest scores.

\begin{table*}[]
\caption{Questions about student general CS self-efficacy level}
\label{selfefficacy}
\centering
\begin{tabular}{l}
\hline
{\textbf{Question Item}}\\ \hline
{1 - Generally I have felt secure about attempting computer programming problems.}  \\
{2 - I am sure I could do advanced work in computer science.}   \\
{3 - I am sure that I can learn programming.} \\
{4 - I think I could handle more difficult programming problems.}     \\
{5 - I can get good grades in computer science.} \\
{6 - I have a lot of self-confidence when it comes to programming.}   \\ \hline
\end{tabular}
\end{table*}

\begin{table*}[]
\caption{Questions and ratings on prior-practice knowledge levels in Python \texttt{class}}
\label{knowledge}
\begin{tabular}{ll}
\hline
\multicolumn{1}{c}{\textbf{Question Item}} & \multicolumn{1}{c}{\textbf{Rate Level}}  \\ \hline
\begin{tabular}[c]{@{}l@{}}
1 - Creating classes like \codeword{class Person:} \\ and \codeword{objects} like \codeword{p = Person("XXX")}\end{tabular} & 1 - I am unfamiliar with these concepts   \\
2 - Methods like \initspace and  \str  & 2 - I know what they mean, but have not used them in a program   \\
3 - The use of \codeword{self} in \codeword{class} & \begin{tabular}[c]{@{}l@{}}3 - I have used these concepts in a program, but am not confident \\  about my ability to use them\end{tabular} \\
\begin{tabular}[c]{@{}l@{}}
4 - Defining instance variables like \\ \codeword{self.color = color}\end{tabular}    & \begin{tabular}[c]{@{}l@{}}4 - I am confident in my ability to use these concepts in simple  programs\end{tabular}  \\  & \begin{tabular}[c]{@{}l@{}}5 - I am confident in my ability to use these concepts in complex programs\end{tabular}                                                                \\ \hline
\end{tabular}
\end{table*}

\subsection{Materials}
There were five parts to the materials: basic knowledge instruction on how to create \codeword{classes, methods, objects} in Python; a survey to measure students' CS self-efficacy level and pre-practice knowledge level about writing \codeword{class}; four write-code practice problems and corresponding equivalent Parsons problems; and an immediate posttest to assess their learning performance. \par 

\subsubsection{Basic knowledge instruction}
\begin{figure}[ht]
    \centering
    \includegraphics[width=\linewidth]{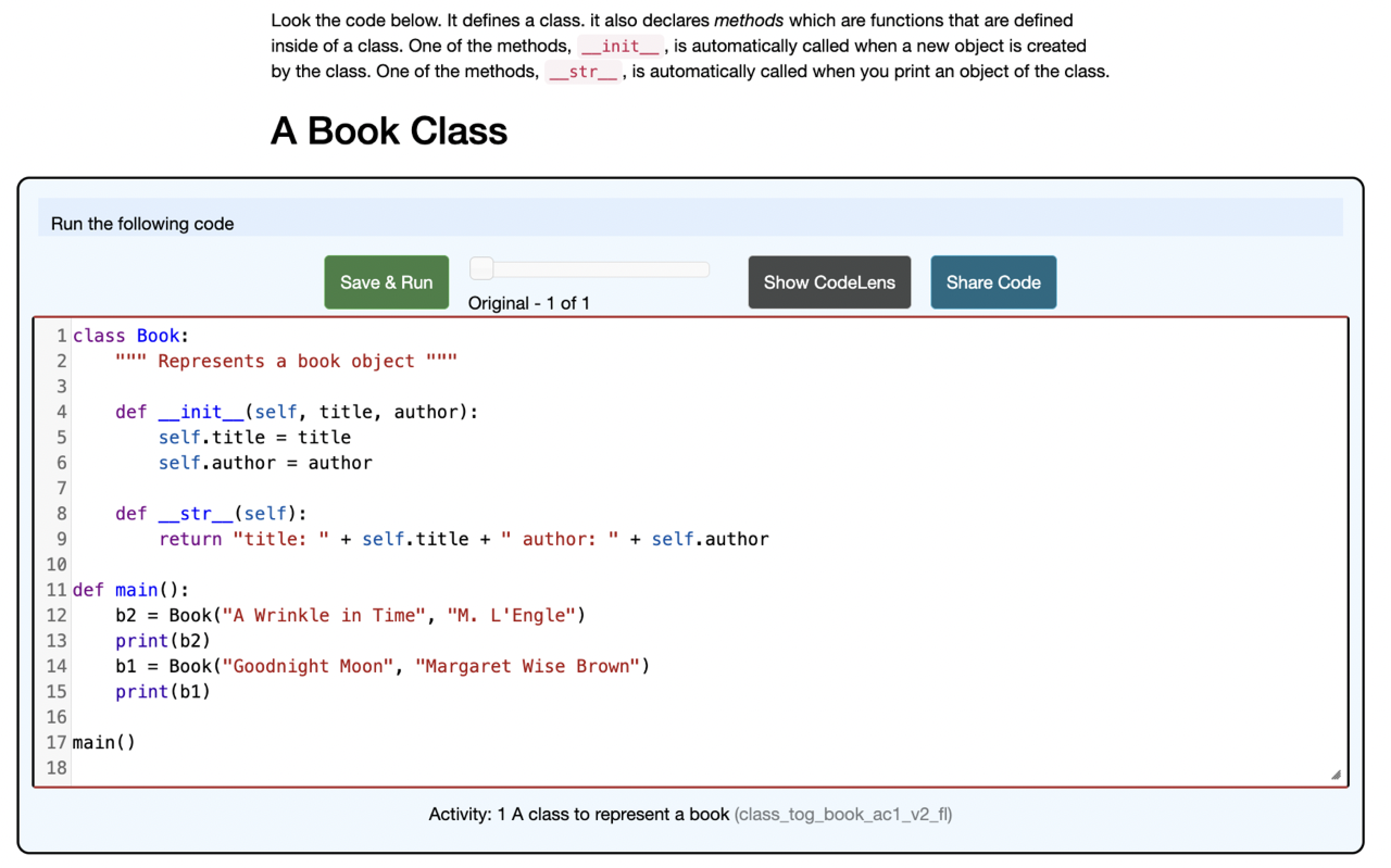} \\
    \caption{Screenshot of the first part of the basic knowledge instruction}
    \label{worked_example}
\end{figure}

The knowledge introduction described how to create a new  \codeword{class}, how to override the inherited \initspace and \str methods, how to define a new \codeword{method}, and how to create \codeword{objects}. For each subtopic, we provided a combination of textual introduction and an interactive worked example that the participant could execute and revise to learn how to handle a specific problem in Python \cite{atkinson2000learning}. Students in this study had little prior experience with creating a new  \codeword{class} in Python, so we believed this basic knowledge instruction part would provide students with some conceptual and procedural knowledge on this topic (Figure \ref{worked_example}).
\vspace{-1mm}
\subsubsection{Survey Items} To gather students' self-efficacy in computer science learning in general (their general CS self-efficacy level), we used the self-efficacy items developed for the computer science domain \cite{wiggins2017you}. The scale to measure self-efficacy was the first 6 items from a 13-item subscale developed by \citeauthor{wiebe2003computer} \cite{wiebe2003computer}. Participants were asked to rate these statements on a five-point scale, ranging from "1-strongly disagree" to "5-strongly agree" (Table \ref{selfefficacy}). This resulted in a score scale of 6 to 30. 

Additionally, to assess whether there was a significant difference between the two groups, we used \citeauthor{duran2019exploring}'s method \cite{duran2019exploring} to collect self-ratings for the pre-practice knowledge level on this topic. Self-evaluation instruments can be used to capture pedagogically useful information about students’ prior programming knowledge \cite{duran2019exploring}. To measure students' pre-practice knowledge level, we created a four-question survey around  \codeword{class} and asked students to rate how well they understood the concepts and could proceed with translating them into actual code (Table \ref{knowledge}). For the rate levels, given that this instrument was used to measure a new topic, we created five different levels adapted from the original prototype self-evaluation instrument (Table \ref{knowledge}), and assigned a score of 1-low to 5-high for each level, with a total score range of 4 to 20. 

\subsubsection{Write-code Practice Problems} We sourced the write-code problems from an intermediate Python programming course at a public research university in the United States. We chose four write-code practice problems, each worth 10 points (Table \ref{practice}). The first problem required students to create a class with an \initspace and \strspace method. The second and third problems involved implementing a third method in addition to \initspace and \str. The fourth problem was the hardest one which required performing a random selection from a list, in addition to creating a class and a method. Our goal was to make sure that participants with different skill levels all had a chance to use a Parsons problem to scaffold a write-code problem. During practice, students from the two conditions both received execution-based feedback after each run, which showed compiler errors and the output from running the code including the results from unit tests (Figure \ref{write}). Each problem was scored out of 10 points based on the percentage of unit tests that passed. There were a total of 40 possible points for the overall practice. In this work, we first extracted the practice results provided by the auto-grading system. Then, we manually checked the final code submissions and accounted for any auto-grading errors before conducting the analyses. \par

\vspace{-2mm}
\subsubsection{Parsons Problems as Scaffolding} One recent study of Parsons problems found that a Parsons problem with an unusual solution increased students' cognitive load \cite{haynes2021problem}. To address this problem, we clustered student-written code from previous semesters using the OverCode software \cite{glassman2015overcode}, and then used the most common student solution cluster to create an equivalent Parsons problem (Figure \ref{interface}). To highlight common misconceptions, we also inserted paired distractors into the Parsons problems. Experts created the distractor blocks based on common syntax or semantic errors \cite{parsons2006parson}. Additionally, a “Help Me” button was provided at the bottom of each Parsons problem to assist students who struggle while solving the Parsons problem. This button triggered intra-problem adaptation, which either removes a distractor or combines two blocks into one if the learner had made at least three attempts to solve the Parsons problem and there are more than three blocks left in the solution.

\subsubsection{Posttest Items} Two types of questions were included in the posttest: write-code problems (10 points each, 20 points in total) and fix-code problems (10 points each, 20 points in total). In a write-code problem, the student needs to write the correct code from scratch following the problem description, and in a fix-code problem, the student must fix the errors in the existing buggy solution. These questions covered similar concepts to the write-code practice questions. Every time they ran the code, students would receive execution-based feedback, which showed compiler errors and the output from running the code, usually including results from unit tests. We calculated their posttest scores by using the proportion of passed unit tests in the final submission. 

\begin{table*}[]
\caption{Four Write-code Practice Problems}
\label{practice}
\begin{tabular}{ll}
\hline \textbf{Problem Name} & \multicolumn{1}{c}{\textbf{Question Item}}                                                                                                     \\ \hline
{1 - \textit{Song}} & Write a class \codeword{Song} with an \initspace method that takes a \codeword{title} as a string and \codeword{len} as a number \\ & and initializes these attributes in the current object. Then define the \strspace method to return \\ & the \codeword{title, len}.  For example,  \codeword{print(s)}  when \codeword{s = Song('Respect',150)} would print \\ &  “Respect, 150”.                                                                     \\
{2 - \textit{Cat}} & Write a class \codeword{Cat} with an \initspace method that takes \codeword{name} as a string and \codeword{age} as a number \\ &  and initializes these attributes in the current object. Next create the \strspace method that returns \\ &  “name: name, age: age”. For example if \codeword{c = Cat("Fluffy", 3)} then \codeword{print(c)} should print \\ & "name: Fluffy, age: 3". Then define the \codeword{make\_sound} method to return "Meow". \\
{3 - \textit{Account}} & Create a class \codeword{Account} with an \initspace method that takes \codeword{id} and \codeword{balance} as numbers. Then \\ & create a \strspace method that returns “id, balance”. Next create a \codeword{deposit} method takes \codeword{amount} \\ & as a number and adds that to the balance. For example, if \codeword{a = Account(32, 100)}  and \\ & \codeword{a.deposit(50)} is executed, \codeword{print(a)} should print “32, 150”.                           \\
{4 - \textit{FortuneTeller}} &  Write a class \codeword{FortuneTeller} with an \initspace method that takes a list of fortunes as strings and \\ & saves that as an attribute. Then create a \codeword{tell\_fortune} method that returns one of the fortunes \\ & in the list at random.                          \\ \hline
\end{tabular}
\end{table*}

\section{Results}

\subsection{RQ1: Are there differences between conditions in terms of practice performance, problem-solving efficiency, and posttest performance for students with low CS self-efficacy levels (RQ1.1) and for students with high CS self-efficacy levels (RQ1.2)?}

Descriptive statistics on student responses to the survey items are included in Table \ref{prepost}. As shown in Table \ref{prepost}, on average, students in both conditions reported a pre-practice knowledge level between "I know what it means, but have not used it in a program" (8 total points) and "I have used this concept in a program, but am not confident about my ability to use it" (12 total points). This demonstrated that, after receiving basic direct instruction with worked examples, these students gained some confidence in their ability to write code around \codeword{class}. However, they have not become experts on this topic. This indicated that those participants reached the expected skill level for this study. In addition, as the conditions (NP and PS) were randomly assigned, we expected students in both conditions to have a similar level of self-rated prior knowledge in Python \codeword{class} and general CS self-efficacy before they started the practice. Given that their self-rated pre-practice knowledge and general CS self-efficacy were not normally distributed, we applied two Mann-Whitney U tests, and our statistical results indicated that there were no significant differences between PS and NP conditions on their basic pre-practice knowledge (\textit{U} = 1008.0, \textit{p} = .846, CLES = .51) or general CS self-efficacy (\textit{U} = 1146.0, \textit{p} = .182, CLES = .58), suggesting that the condition groups were comparable. We also used Cronbach’s \(\alpha\) reliability test to check the internal consistency of the survey questions, resulting in \(\alpha\) = 0.82 for the general CS self-efficacy survey and  \(\alpha\) = 0.80 for the pre-practice knowledge in Python \codeword{class} survey, demonstrating that these two surveys had good internal consistency \cite{gliem2003calculating}.

\begin{table*}[]
\caption{Pre-practice knowledge level and general CS self-efficacy level by condition, reported in \textit{M (SD)}, \textit{Mdn} (25th percentile - 75th percentile) format}
\label{prepost}
\begin{tabular}{ccc}
\hline
\textbf{Category (Score range)}                        & \textbf{PS} (\textit{N}=41)       & \textbf{NP}  (\textit{N}=48)            \\ \hline
pre-practice knowledge in Python \codeword{class} (4-20) & 9.3 (4.9), 8.0 (4.0-13.0)  & 8.8 (4.3), 7.5 (5.8-12.0) \\
general CS self-efficacy level (6-30) & 21.1 (4.8), 21.0 (18.0-24.0) &  19.7 (4.6), 19.0 (17.0-23.0) \\
\hline
\end{tabular}
\end{table*}

We calculated each student's in-practice problem-solving efficiency using the likelihood model from \citeauthor{hoffman2010conceptions} \cite{hoffman2010conceptions}. A learner would have a bigger relative gain and be seen as more efficient if they spent less time while still achieving more problem-solving accuracy \cite{hoffman2010conceptions}. Following this model, for each student, the in-practice problem-solving efficiency was calculated as the ratio of practice score (\textit{Max} = 40) and practice time (mins). Practice time was calculated as the time used for practice, excluding any periods of inactivity over 5 minutes. The highest problem-solving efficiency was 4.07, achieved by a student who finished all four write-code problems (40 points) in 9.83 minutes.

To investigate how Parsons scaffolding may impact learners with varying CS self-efficacy levels differently, we divided learners into two groups based on their scores in the general CS self-efficacy survey. The PS-High (\textit{N} = 20, \textit{M} = 25.1, \textit{SD} = 2.7) and NP-High (\textit{N} = 22, \textit{M} = 23.8, \textit{SD} = 2.7) are students that scored higher on the CS self-efficacy survey; the PS-Low (\textit{N} = 21, \textit{M} = 17.3, \textit{SD} = 2.8) and NP-Low (\textit{N} = 26, \textit{M} = 16.2, \textit{SD} = 2.4) are students that scored lower in the general CS self-efficacy survey.

We found no differences between PS-High \& NP-High groups in terms of the self-evaluated prior knowledge in Python \codeword{class} (\textit{U} = 230.0, \textit{p} = .810, CLES = .52) and PS-Low \& NP-Low groups (\textit{U} = 275.5, \textit{p} = .965, CLES = .50). We then conducted a series of analyses to understand the differences in terms of practice performance, problem-solving efficiency, and posttest performance among the groups. In cases where the data was not normally distributed, we used the Mann–Whitney U test instead of ANOVA. Results are included in Table \ref{RQ1}. We observed significant differences between PS-Low and NP-Low groups on write-code practice performance and problem-solving efficiency; PS-Low students had significantly higher write-code practice performance and problem-solving efficiency than NP-Low students. However, there is no significant difference in write-code practice performance and problem-solving efficiency between students in PS-High and NP-High groups. In addition, there are no significant differences in posttest performance between the two conditions for students in both low and high CS self-efficacy groups.

\begin{table*}[]
\caption{Comparison of practice performance, problem-solving efficiency, and posttest performance by condition at high and low CS self-efficacy levels, reported in \textit{Mdn} (25th percentile - 75th percentile) format }

\label{RQ1}
\begin{tabular}{p{0.222\textwidth}p{0.245\textwidth}p{0.217\textwidth}p{0.25\textwidth}}
\hline
\textbf{Category}       & \textbf{Parsons Scaffolding - High} & \textbf{No Scaffolding - High} & \textbf{Statistical Results}    \\ \hline
Write-code practice   & 20.0 (10.0-40.0)      & 22.5 (0.0-35.0) & \textit{U} = 254.5, \textit{p} = .381, CLES = .58  \\
Problem-solving efficiency& 1.5 (0.7-2.4)     & 1.2 (0-2.7)   & \textit{U} = 242.0, \textit{p} = .585, CLES = .55 \\ 
Posttest performance  & 10.0 (0-36.7)  & 25.0 (0-36.2) & \textit{U} = 211.5, \textit{p} = .837, CLES = .48 \\ \hline
\textbf{Category}       & \textbf{Parsons Scaffolding - Low} & \textbf{No Scaffolding - Low} & \textbf{Statistical Results}       \\ \hline
Write-code practice  & 20.0 (10.0-40.0)   & 0 (0-0)  & \textit{U} = 442.5, \textit{p} < .001, CLES = .81  \\
Problem-solving efficiency & 1.3 (0.5-1.7)  & 0 (0-0) & \textit{U} = 435.0, \textit{p} < .001, CLES = .80  \\
Posttest performance & 0 (0-30.0)     & 0 (0-5.3) & \textit{U} = 335.0, \textit{p} = .138, CLES = .61  \\

\hline
\end{tabular}
\vspace{-2mm}
\end{table*}

\subsection{RQ2: In the Parsons Problems as Scaffolding (PS) condition, how did students with varying CS self-efficacy levels use the Parsons scaffolding?}

In order to answer this RQ, we first need to describe the expected behavior of utilizing Parsons problems as scaffolding. Since the Parsons problem is optional and the scaffolding is supposed to be initiated by the students, we did not expect every student to use the Parsons problem to solve every write-code problem; we wanted them to use it when they were in need of help. In addition, as previously described in Section \ref{scaffolding-section} and Section \ref{Method}, students had the autonomy to engage with the Parsons problem in any way they chose. However, the different ways were interdependent because students were only able to interact with Parsons scaffolding in one specific way for each question. Therefore, to avoid redundant analyses, we selected the most extensive use of Parsons scaffolding (solve) to get a sense of the relationship between students' overall CS self-efficacy and their utilization of Parsons scaffolding. We calculated \textit{solve Parsons scaffolding rate} as the total number of times they solved Parsons scaffolding divided by the number of practice problems, which is four in our study. The mean (\textit{M}) was 35\%, with a standard deviation (\textit{SD}) of 38\%, ranging from 0\% to 100\%.  We then computed a Pearson correlation and found a significant negative correlation between students' general CS self-efficacy and solve Parsons scaffolding rate, \textit{r} = -.32, \textit{p} = .043. Specifically, when given Parsons problems as scaffolding during practice, students with lower CS self-efficacy were more likely to solve them. In other words, students with higher CS self-efficacy tended to use Parsons scaffolding more lightly, or even solve problems independently.


Given that the ultimate goal of providing scaffolding is to help students solve write-code problems, we are also interested in knowing the relationship between their CS self-efficacy levels and how they used Parsons scaffolding to solve write-code problems. As a result, for each student, we then computed the \textit{effective scaffolding rate} as the frequency of using the Parsons scaffolding in any of the three methods and completing the corresponding write-code problem divided by the number of times they used the Parsons scaffolding. A high effective scaffolding rate (close to 1) indicates that the scaffolding was more effective in helping the students finish the write-code practice, while a low rate (close to 0) indicates the student did not benefit from the scaffolding as much. For instance, if a student used the Parsons problem for three write-code practice problems but only finished one write-code problem after using the scaffolding, then the effective scaffolding rate would be 0.33. The mean and standard deviation of the effective scaffolding rate is \textit{M} = 51\%, \textit{SD} = 40\%. We then computed the Pearson correlation between the effective scaffolding rate and students' general CS self-efficacy scales, which is \textit{r} = .18, \textit{p} = .292. 

\textbf{Posthoc Analysis} Considering that the average effective scaffolding rate is 51\%, this suggests that some Parsons scaffolding usage did not reach the expected outcome in helping students solve the write code problem. To better understand why and how some PS students used Parsons scaffolding less effectively, we conducted two follow-up analyses. We firstly computed the Pearson correlation between students' effective scaffolding rate and students' pre-practice knowledge level on Python \codeword{class}, and observed a significant positive relationship, \textit{r} = .40, \textit{p} = .014. 

In addition, for those who utilized the Parsons scaffolding but still got the write-code practice wrong, we looked deeper into their final write-code submissions and the corresponding final Parsons scaffolding problem status. This resulted in a total of 47 Parsons scaffolding state \& write-code state pairs. We classified these pairs into four categories based on the final Parsons scaffolding state, followed by the corresponding final code submission state:
(1) \textit{Only scanned Parsons blocks} (26 instances) - twenty-four (92\%) of their final write code submissions did not compile due to syntax errors, type errors, or name errors, one student did not write any code after viewing Parsons blocks, and one student's code compiled successfully but had logic errors and did not pass all the unit tests.
(2) \textit{Completed Parsons problems but failed write-code task}  (15 instances) - twelve of them had code in the write-code box, and three left the write-code box blank. In all 12 cases, students did the final write-code submission after getting the Parsons solution but still failed to solve it. Since they already had the correct solution in hand (the Parsons solution), we examined their actual code to understand the specific errors they made better. Our results showed that, while these students completed the Parsons scaffolding problem, they still might not have acquired enough skill in writing code on this topic. For example, in six cases, students received a type error by writing \codeword{\_init\_} or \codeword{\_str\_}, which should be \init or \str. And two students had syntax errors in the longest line of \textit{Song}: \codeword{return self.title + ", " + str(self.len)}, such as missing the \codeword{+} or miswriting as \codeword{str().self.len}. We also found one student incorrectly placed the correct solution after the default test cases for three problems. (3) \textit{Incorrect Parsons completion} (2 instances) - both of them omitted some problem requirements and did not finish the code. For example, one student did not complete the required \codeword{def deposit(self,amount)} by leaving this part blank; (4) \textit{Attempted to move Parsons blocks but could not finish} (4 instances) - two final code submissions did not complete the requirements, and two did not write any code. 

In summary, we found that students with lower levels of general CS self-efficacy were more likely to solve Parsons problems as scaffolding during practice. Furthermore, when choosing to use Parsons scaffolding, students with higher pre-practice knowledge in Python \codeword{class} were more likely to use it effectively. However, the effectiveness was not related to students' CS self-efficacy levels. Our preliminary analysis of students' code submissions revealed that the current Parsons scaffolding method might still be too challenging for some students.

\vspace{-3mm}
\subsection{RQ3: In the Parsons problems as Scaffolding (PS) condition, how did students rate the usefulness of the Parsons scaffolding and why?}

\begin{figure}[ht]
    \centering
    \includegraphics[width=\linewidth]{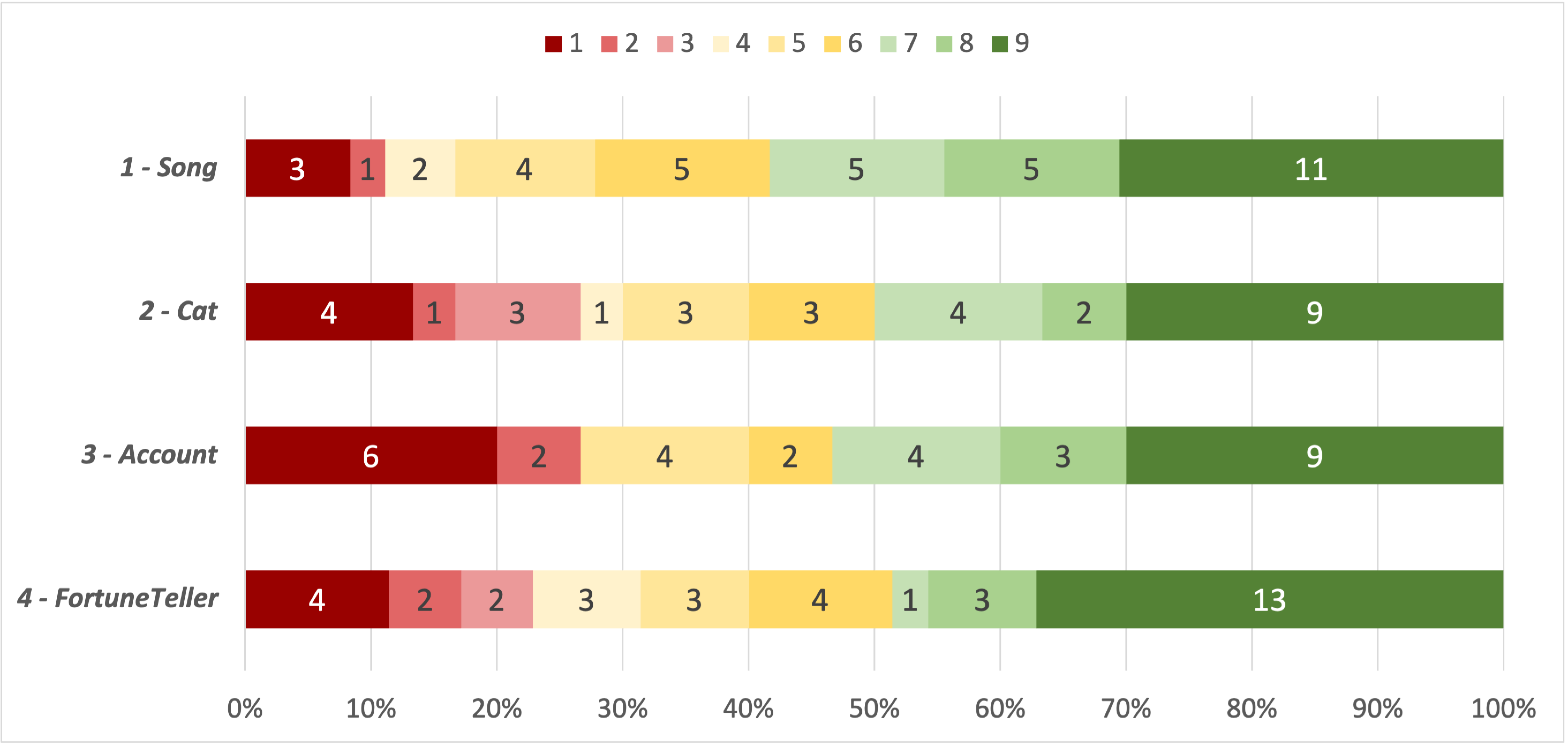} \\
    \caption{Stacked bar chart of Parsons scaffolding usefulness ratings for each practice problem}
    \label{rating}
    \vspace{-3mm}
\end{figure}

To get a general sense of how useful this scaffolding method was, we first looked at students' ratings on the usefulness of the Parsons problems as scaffolding. If students did not apply the scaffolding for the write-code question, they were told to skip the related rating question. The distribution of student ratings for each problem is shown in Figure \ref{rating}. We found that for all four problems, nearly 70\% of the ratings were five or more. This indicates that students generally found Parsons problems helpful in solving the write-code practice problems, which target new concepts they are learning. In addition, 71\% (29 out of 41) of students in the PS condition completed the open-ended question to explain their ratings. Specifically, 19 (66\%) of them explained how Parsons scaffolding helped them while two of them reported specific challenges they faced, and the rest of the eight answers (28\%) explained their ratings by discussing how difficult this practice was for them.

The average usefulness rating for those 19 students was 7.7, indicating that those students found Parsons scaffolding to be helpful. Of those 19 positive explanations, two responses only stated that Parsons problems were helpful without further elaboration. For the rest of the answers, four students (21\%) stated that Parsons problem made it easier for them to solve the write-code problem, but still learn. For example, one student wrote \textit{"The Parsons problem gives me all the necessary elements of creating a} \codeword{class} \textit{which I am unfamiliar with, but I still have to figure out an order which is helping me learn, but not too strenuous".} Four students (21\%) thought Parsons problem helped them learn problem-solving strategies, as one student explained, \textit{"I feel like Parsons helped me understand the different pieces of code and how to think about the problems."} A high proportion of students (47.38\%, 9 out of 19 students) valued the benefit of refining and extending existing programming knowledge by using Parsons scaffolding. One student claimed that \textit{"I am still confused about} \codeword{class} \textit{,} \codeword{method} \textit{, and} \codeword{self}, \textit{but these problems definitely helped me understand better"}, another one expressed a similar idea, \textit{"the Parsons problems helped me remember how to create} \codeword{methods} \textit{and enter arguments as necessary."} Additionally, one student elaborated on how the Parsons scaffolding helped to extend the existing programming knowledge in a more detailed way, \textit{"I understood structurally what I had to do; I just forgot how to return one of the fortunes in the list at random, so the Parsons problem helped me figure out how to accomplish that."} 

However, some participants reported difficulties when using Parsons scaffolding to complete write-code problems, as evidenced by their low usefulness ratings and corresponding negative explanations. For instance, the student who rated \textit{Cat} three and \textit{FortuneTeller} one did not finish those two problems and explained the low ratings further as \textit{"I still do not understand} \codeword{class} \textit{and found this exercise extremely frustrating ..."} Similarly, the student who rated \textit{Account} one only scanned the Parsons problem and did not finish the write-code problem. This student emphasized the inherent high difficulty as \textit{"This was really hard to just dive right into."} As mentioned above, we also had two students who explicitly explained why Parsons scaffolding was less useful for them. One of the two students used the Parsons scaffolding four times but was unable to finish any of the associated write-code problems. This student gave an average usefulness rating of 4.8 and provided the following explanation: \textit{"I think it helps me get started, but because this is completely new to me, it merely helps me see what is there, not how to put it together."} Another student received an effective scaffolding rate of 0.5 and reported an average usefulness rating of 4.3. This student explained the rating further as \textit{"... Parsons isn't really that helpful because you're just copying. "}

\section{Discussion}
This study investigates whether and how using Parsons problems can scaffold the write-code practice with students of different CS self-efficacy levels. We found that, for newly learned difficult programming concepts, when having Parsons problem as write-code scaffolding, students with low CS self-efficacy (PS-Low) achieved significantly higher practice performance and problem-solving efficiency than those with low CS self-efficacy in the no scaffolding condition (NP-Low). However, this condition effect did not occur in students with high CS self-efficacy (PS-High and NP-High). Also, there were no condition effects on posttest performance among the groups.

Beyond intervention effects, for PS students, we also examined the relationship between their general CS self-efficacy and the use of Parsons scaffolding. We discovered that students with lower levels of CS self-efficacy are more likely to solve Parsons problems as scaffolding during practice. To better understand the ineffective scaffolding use, we also conducted two follow-up analyses and found that, when choosing to use the Parsons scaffolding, students with higher pre-practice knowledge in Python \codeword{class} were more likely to use it effectively; however, this was not related to students' general CS self-efficacy. Further, our preliminary analysis of student code found that this Parsons scaffolding method might still be too challenging for some students. After analyzing students' useful ratings and explanations, we found Parsons scaffolding could help students to learn to write code, but there are still areas to improve. In this section, we will discuss our findings regarding the condition effects for students with high and low CS self-efficacy levels (RQ1). In the next section, we will present some design implications based on the results of RQ2 and RQ3.
\subsection{RQ1: Parsons scaffolding is more beneficial for students with low CS self-efficacy in terms of practice performance and in-practice problem-solving efficiency}

According to our results, for students with low general CS self-efficacy, those who received Parsons scaffolding (PS-Low) achieved significantly higher practice performance and in-practice problem-solving efficiency compared to those who did not receive scaffolding (NP-Low). However, we did not obtain such condition effects between students with high CS self-efficacy (PS-High and NP-High). This could be explained by the idea that a person with a higher level of self-efficacy will exert more effort on learning tasks and will persevere when confronted with difficulties \cite{britner2006sources, zeldin2000against,wang2022examining}. Given that the programming concept we used in this study was relatively new to students, most of them had little prior experience writing code about it except for the three examples of code they could execute in the instruction phase. However, students with high self-efficacy in computer science, regardless of whether they received Parsons scaffolding or not, were likely more willing to continue with this frustrating process of writing code. Therefore, they achieved a similar level of practice performance and problem-solving efficiency during practice. 

On the other hand, when there was no scaffolding, solving those write-code problems may seem hopeless for students with low CS self-efficacy. Therefore, it was common for NP-Low students to give up quickly rather than persevere in overcoming the difficulties. Nevertheless, although PS-Low students were also prone to giving up, Parsons problems were there to assist them in solving write-code problems that they would not have been able to solve on their own. Consequently, by providing a more supportive learning experience, students with lower CS self-efficacy could finish more write-code practice and achieve higher in-practice problem-solving efficiency.

As for posttest performance, we found no significant differences among the groups. This outcome can be explained by the inherent high difficulty of writing code for a newly learned concept from scratch. Although PS students practiced more successfully than NP students during practice in general, it may not have been enough for them to become proficient in code writing with \codeword{class} because of the small number of practice problems we applied in this study \cite{atkinson2000learning}. In the future, we could use student models to track their write-code skill development and determine the number of practice problems they require to master this topic  \cite{corbett1994knowledge}.


\subsection{RQ2 \& RQ3: Improve the effectiveness of using Parsons problems to scaffold write-code exercise}
Our result from RQ2 indicated that students with lower CS self-efficacy were more likely to solve the Parsons problem when they had the scaffolding. One possible explanation is that learners with lower CS self-efficacy were less confident in their ability to perform the write-code problem successfully, similar to \citeauthor{wang2022examining} \cite{wang2022examining}. As a result, they were more willing to solve a partially correct solution (Parsons problem) and follow it. However, students with higher CS self-efficacy levels were more confident in completing the write-code exercise. They would rather spend more time on their own code and, if necessary, use Parsons scaffolding in a more limited way. In addition, we also found that not all the Parsons scaffolding usage was effective. We found that students with a lower pre-practice knowledge level are more likely to use Parsons scaffolding ineffectually. Our analysis of their code also revealed several ineffective scenarios, which led to design suggestions for different groups of students.


Firstly, one possible reason why some students struggle with using Parsons scaffolding is that it may still be too difficult. Although we implemented a "Help Me" button for student-initiated difficulty level adaptation, students still had to complete at least three full attempts to activate it. It was possible that students became overwhelmed by the default multiple Parsons blocks, causing them to give up before even attempting three times. Therefore, an important enhancement of the current scaffolding approach would be providing the Parsons problems with appropriate levels of difficulty based on students' submission histories or incorrect code. In addition, to prevent students from feeling overwhelmed by encountering a new-looking problem (the equivalent Parsons problem) other than the write-code problem, it is important to establish a connection between the Parsons scaffolding and their existing code. One way to achieve this is to personalize the Parsons problem. Instead of providing the most common previous student solution, we could create a Parsons problem that is based on the student's incorrect code. Furthermore, to give students a sense of accomplishment and relatedness between existing written code and Parsons scaffolding, we could make the correct parts of the existing student code static with positive feedback \cite{marwan2020adaptive}.

Secondly, we found that some students finished the Parsons problem as scaffolding but still had errors in their write-code submissions. One potential reason is that, while these students finished the Parsons scaffolding problem, they were unable to localize their errors, such as the double leading and trailing underscores in Python, and could not use the Parsons solution effectively. Alternatively, students might prefer not to follow the Parsons solution, and therefore keep the errors in their own approach. These call for a more personalized Parsons scaffolding approach, which includes using solutions close to the students' current path and adding paired distractors that can highlight errors based on the student's current code state. Besides refining the scaffolding mechanism, incorporating explanations is another way to boost the efficiency of Parsons scaffolding. One direction to achieve this is to provide prompt guidance to make learners' self-explanation on Parsons scaffoldings more productive since it helps them concentrate on relevant information \cite{margulieux2019finding, nguyen2023examining}. Another direction is adding textual explanations to either Parsons blocks or the finished Parsons solution. A prior study by \citeauthor{marwan2019impact} added explanations to next-step hints and found that novices thought hints with explanations were much more relevant and understandable \cite{marwan2019impact}. They were also better able to relate these hints to their code. We believe this method could improve students' effective scaffolding rate.

Moreover, given that we found some students finished the Parsons scaffolding and left the write-code box blank, it is possible that those students felt it was unnecessary to retype the correct Parsons solution into the write-code box. While we think that the process of retyping might enrich their comprehension and help them to identify some points that they might not have noticed without typing on their own, we could add an "AutoFill" button for those who made considerable progress when utilizing the Parsons scaffolding. 


\section{Limitation and future work}
One limitation of this work is that we did not collect subjective ratings or perceptions from participants who did not use Parsons problems as scaffolding. We plan to add some concrete survey questions to learn more about why they chose not to use the scaffolding. In addition, future retrospective interviews are necessary to fully understand the ineffective scaffolding cases. Moreover, we only conducted this study on one topic in a medium-scale classroom at one public university in the United States. With other demographic groups, computing domains, and educational settings, like data science and MOOCs, we might observe other Parsons usage scenarios and scaffolding effects. Furthermore, in order to save class time and reduce pretest cognitive overload for students, we only utilized self-reported measures to assess their pre-practice knowledge level in Python \codeword{class}, which may be subject to bias.

Regarding future work, we would like to investigate the impact of Parsons problems as scaffolding with other scaffolding techniques, in other programming languages, and educational settings. In addition, we are excited to continue our work based on the design suggestions provided above, such as providing personalized Parsons problems that are based on the student's existing incorrect solution and adapting the Parsons problem scaffolding by setting the starting state of the Parsons problem as the final state of the student's code. In addition, we also look forward to reducing the cost of developing Parsons problems by automating the process based on student submissions and large language models.
\section{Conclusion}
In this work, we investigated the effects of Parsons problems as scaffolding during code writing skill acquisition for students of various CS self-efficacy levels. We found that lower CS self-efficacy students in the Parsons as scaffolding condition achieved significantly higher practice performance and problem-solving efficiency than lower CS self-efficacy students in the condition without any scaffolding. Further investigations into student interaction with the Parsons scaffolding revealed that students with lower levels of CS self-efficacy are more likely to solve Parsons scaffolding problems during practice. In addition, when choosing to use Parsons problems as scaffolding, students with higher pre-practice knowledge of the topic were more likely to use them effectively; however, this was not related to students' general CS self-efficacy. These findings direct us to optimize the Parsons scaffolding experience, including providing personalized and adaptive versions of the Parsons problems based on the student's current problem-solving status.
\begin{acks}
The funding for this research came from the National Science Foundation award 2143028. Any opinions, findings, and conclusions or recommendations expressed in this material are those of the authors and do not necessarily reflect the views of the National Science Foundation.
\end{acks}

\bibliographystyle{ACM-Reference-Format}
\bibliography{bibliography}

\end{document}